\documentclass{aa}
\usepackage{graphicx}
\usepackage{subfigure}
\usepackage{natbib}

\begin{document}

\title{Critical review of the results of the Homestake solar neutrino experiment}
 
\author{Paolo Walter Cattaneo} 

\institute{INFN Sezione di Pavia and Dipartimento di Fisica Nucleare e 
Teorica, Via Bassi 6, Pavia, I-27100, Italy\\
\email{Paolo.Cattaneo@pv.infn.it}}

\date{Received November 30, 2002; accepted November 30, 2002}

\abstract{
The radiochemical experiment in the Homestake mine was designed to measure the
solar neutrino flux through the detection of $^{37}Ar$ produced in the reaction 
$\nu_e + ^{37}Cl \longrightarrow e^-+^{37}Ar$. The comparison between this 
measurement and the theoretical predictions from solar models evidences a 
substantial disagreement.
I reanalyzed the data evidencing a bias with high statistical significance and
suggesting a new interpretation of the data.
}

\titlerunning{Homestake experiment critical review}
\maketitle

\section{Introduction}

The Homestake chlorine experiment has been running for over 20 years 
providing measurements of a portion of the solar neutrino flux. A detailed 
description of the experimental apparatus and of the analysis is given in
\cite{homestake} and \cite{neutastro}.\\
Briefly, the experiment consists of about 133 tons of $^{37}Cl$ in the form of 
$C_2Cl_4$ located in a tank in the Homestake mine. The solar neutrinos induce 
the reaction $\nu_e + ^{37}Cl \longrightarrow e^-+^{37}Ar$ and the resulting 
$^{37}Ar$ is extracted and put into proportional counters, that measure energy and
timing of each decay. The $^{37}Ar$ atoms are counted observing the 
$2.82\,\mathrm{keV}$ Auger electrons from the electron capture with a half life
of $35.04 \mathrm{d}$. The Auger electrons are selected by appropriate cuts on 
the rise time and selecting an energy window around the peak.\\
A run results in a time series of decays that is fit to the exponential
decay of $^{37}Ar$ plus a decaying background. The $^{37}Ar$ production rate and
the background level in each run are obtained by maximizing the probability of 
obtaining the given time series with a maximum likelihood technique optimized for 
low counting rate \cite{cleveland}. The same fit gives the "$1\sigma$" errors on 
rate and background interpreted as $68\%$ confidence range. The results are 
presented separately for each run, that covers approximately two months of data 
taking.\\
In \cite{homestake} the data are presented as in the past (see \cite{neutastro}) 
in a list of single run analysis obtained using tight cuts on the rise time and 
on the energy window to optimize the dominant statistical error. Alternatively
the data are selected with loose cuts on the rise time and on the energy window
to reduce the systematic errors and a global likelihood analysis is applied to 
all of them. The final measurement of the neutrino flux is based on this latter
analysis and therefore on data that are not presented in detail. Nevertheless, as 
several papers in the past \cite{bafipr}-\cite{bieber}-\cite{krauss}-
\cite{filvogel}-\cite{morrison} and references therein, the single run results 
will be reanalyzed to verify their consistency with the hypothesis of constant 
flux.\\
This hypothesis has been challenged in the past by claims that the flux is 
correlated with the sun spot number or other parameter of solar activity or 
noticing that the fluxes measured over different time intervals differ 
significantly.\\ 
I performed the hypothesis testing of constant flux by calculating the $\chi^2$
of the distribution after having estimated the run errors. That results in a very 
poor agreement with the original hypothesis.\\
Error independent analysises were also performed using the Smirnov and Kolmogorov
hypothesis tests and using rank order statistic analysises.\\
These show the signal and the background levels to be strongly correlated.\\ 
Furthermore the average fluxes and the hypothesis tests are calculated 
separately for different partitions of the run set according to the background 
level rank. The result strongly suggests a bias in the data and a very bad 
$\chi^2$ for the high background subset.\\
Based on this analysis, I give an alternative higher estimation of the flux.

\section{Data and errors}

\begin{figure*}
\includegraphics[width=\textwidth,bb=0 0 567 280]{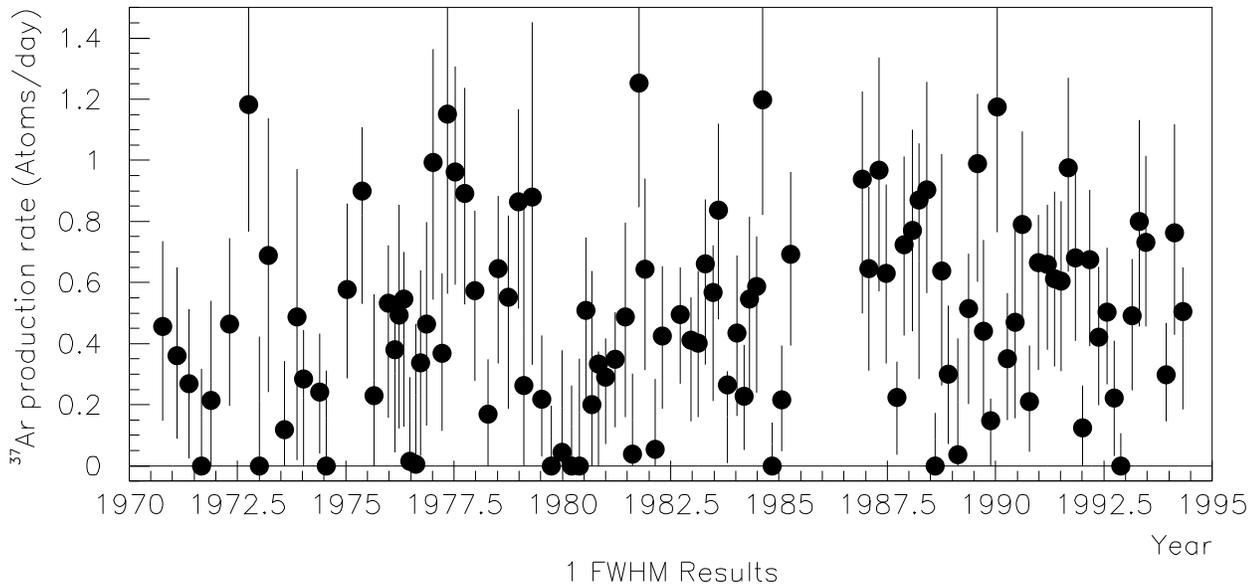}
\caption{Solar neutrino rate with the Homestake experiment on a run basis.}
\label{HScount}
\end{figure*}

\begin{figure*}
\centering
\includegraphics[width=\textwidth,bb=0 0 567 280]{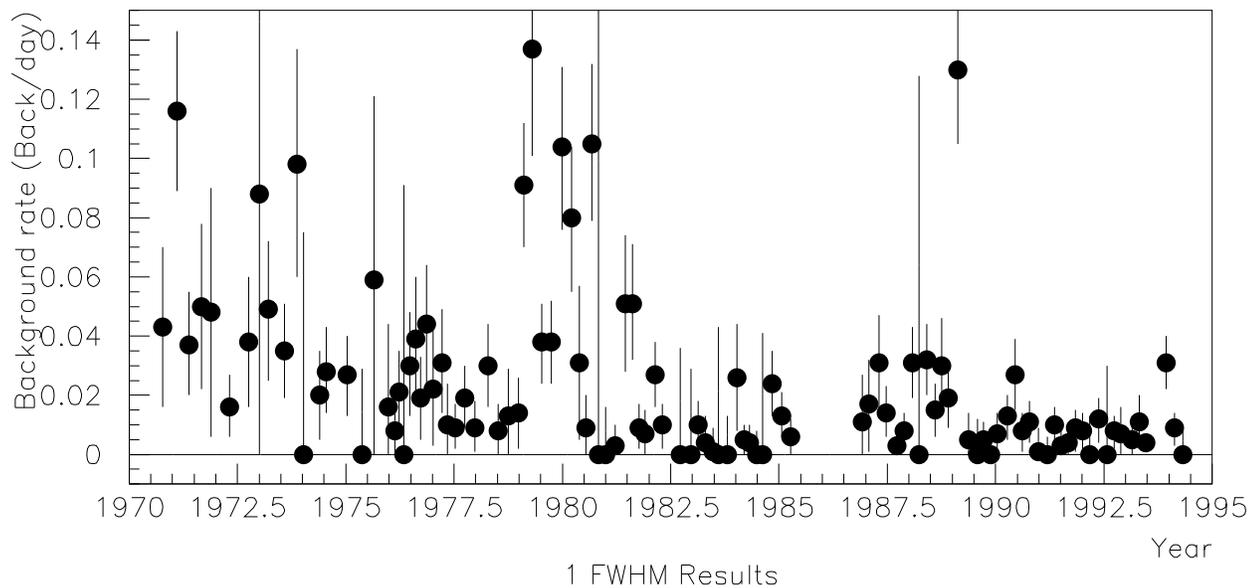}
\caption{Background in the Homestake chlorine experiment on a run basis.}
\label{HSback}
\end{figure*}

The data analyzed in this paper are presented in \cite{homestake} for $N_T=108$
runs in the following format: run start time, run stop time, run average time 
(accounting for the decay of $^{37}Ar$ in the detector), production rate of 
$^{37}Ar$ resulting from the fit 
in atom per day, lower "$1\sigma$" error (68\%\ confidence range) on the 
production rate, higher "$1\sigma$" error (68\%\ confidence range) on the 
production rate, counter background resulting from the fit in count per day, lower
"$1\sigma$" error (68\%\ confidence range) on the counter background, higher 
"$1\sigma$" error (68\%\ confidence range) on the counter background. These data 
are presented in Fig.\ref{HScount} for the production rate and in Fig.\ref{HSback}
for the background.\\
In testing an hypothesis (for example production rate $p$ constant), it is 
necessary to assign an error to the rate measurement of each run. Previous 
analysis have devised different estimation of errors: equal on all runs, implicit 
in using rank-order statistic \cite{bapr}, average of lower and higher errors 
\cite{bafipr}, the larger of the two \cite{neutastro}-\cite{bieber}-\cite{krauss},
calculated by rate and rate errors \cite{filvogel}. All these estimations seems 
incorrect.\\
It is just the case to recall that if a measurement of a Poisson process of average
$\mu$ is $n$, the error is $\sqrt{\mu}$ and not $\sqrt{n}$. This last estimation is
approximately correct only for large $n$, that is in gaussian approximation. If 
only a single measurement is available, the best estimation of $\mu$ is $n$ and 
the two approaches coincide, but if several measurements are available, giving an 
estimation $\tilde \mu$ of $\mu$, the best estimation of the errors on the single
measurement is $\sqrt{\tilde \mu}$.\\
Furthermore, the combination of a Poisson process of average $\mu$ and a binomial 
distribution due to an efficiency $\epsilon$ is a Poisson process of average 
$\mu \epsilon$.
The errors on the production rate $p$ reported by the fit refer to the estimation 
of $p$ in the single run, while the best estimation $\tilde p$ makes use of all 
runs. Therefore, in absence of background counts, zero non-solar neutrino 
production rate and equal efficiency $\epsilon$ for each run, the average counts 
expected in run $i$ of duration $\Delta t_i$ are $n_i={\tilde p}\epsilon \Delta 
t_i^{eff}$, where $\Delta t_i^{eff} = (1-\exp(-\frac{\Delta t_i} {\tau_{^{37}
Ar}}))\tau_{^{37}Ar}$ is the effective run time accounting for $^{37}Ar$ decay. 
Its error is $\sqrt{n_i}=\sqrt{\tilde p \epsilon \Delta t_i^{eff}}$ that gives
an error on the rate $\sigma(\tilde p) = \sqrt{\frac{\tilde p}{\epsilon \Delta 
t_i^{eff}}}$. \cite{cleveland} reports that the efficiency is constant and equal 
to $\epsilon=0.95\times 0.96\times 0.327=0.298$.\\
If the non-solar neutrino rate is non-zero, the Poisson process under measurement
is the sum of two distinct Poisson processes, the solar neutrino rate, $p_S$, and 
the non-solar neutrino rate $p_{i,NS}$, that is supposed known. The measurement 
gives an estimation ${\tilde p_i} = {\tilde p_S}+p_{i,NS}$, from which 
${\tilde p_S} = {\tilde p}-p_{NS}$ (the true known value of $p_{NS}$ is used). 
That gives 
$\sigma_i^2(p_S) = \sigma^2(\tilde p) + \sigma_i^2(p_{NS}) = \sigma^2(\tilde p_S) 
+ 2\sigma_i^2(p_{NS})$, that is $\sigma_i(\tilde p_S) = \sqrt{\frac{\tilde 
p_S+2p_{i,NS}} {\epsilon \Delta t_i^{eff}}}$.\\
In presence of background, the signal can be identified from the decay time of the
atoms. Following a hint in \cite{neutastro}, the background measurement stems 
from the counting in the signal free region at time larger compared to 
$\tau_{^{37}Ar}$, while the signal is measured from the counting within a few
$\tau_{^{37}Ar}$. The statistical effect of the background can be evaluated
expressing the background in term equivalent to signal rate and then applying the
same approach used for the non solar rate component. The background induced rate
is $\tilde p_{bk} = \frac{b_i \tau_{^{37}Ar}}{\epsilon \Delta t^{eff}_i}$, where 
$b_i$ is the background rate. The underlying assumption is that the contribution
of the background to the signal measurement is concentrated in $\tau_{^{37}Ar}$.\\
With these definitions the total error on the solar rate is $\sigma_i(\tilde p_S)
= \sqrt{\frac{\tilde p_S+2p_{i,NS}+2\tilde p_{i,bk}} {\epsilon \Delta 
t_i^{eff}}}$.\\
It is apparent that, in order to use this errors to calculate anything relevant,
an estimation of the rate must be already available. That is obtained through
the unweighted average (actually weighted only through the effective run time
length)
\begin{equation}
\tilde p_{uw} = \frac{\sum_i p_i \Delta t_i^{eff}}{\sum_i \Delta t_i^{eff}}. 
\end{equation}
The weighted average is instead obtained as 
\begin{equation}
\tilde p = \frac{\sum_i \frac{p_i \Delta t_i^{eff}}{\sigma_i^2(\tilde p)}}
{\sum_i \frac{\Delta t_i^{eff}}{\sigma_i^2(\tilde p)}}. 
\end{equation}

\section{Hypothesis testing}

The hypothesis of constant flux or, more generally, of consistency of the data 
set can be tested in several ways \cite{frodesen}-\cite{eadie}.\\
The Pearson's $\chi^2$ method quantifies the consistency of the constant flux 
hypothesis making use of the errors on the full sample, calculating 
\begin{equation}
\chi^2(N-1) = \sum_i \left(\frac{p_i - \tilde p}{\sigma(p_i)}\right)^2
\end{equation}
where N is the number of data points.\\
Partitioning the data set in two subsamples A and B, the same approach can be 
repeated on each subsample as long as the estimator $\tilde p$ is replaced by the
subsample estimator $\tilde p^{A(B)}$. A test of compatibility between the two 
samples is obtained estimating the probability of the difference between the two
estimators.\\ 
The disadvantage of this approach is that it strongly relies on the exact 
evaluation of the errors. As discussed previously the presence of background 
creates some ambiguity in defining them.\\
Alternatively we can assume that all the data have the same weights and employ 
the hypothesis test of Kolmogorov and Smirnov. The test requires to order the 
N observation on the variable $p$ in ascending order ($x_1 \cdots x_N$) and 
build their cumulative distribution 
\[
S_N(x) = \left\{ \begin{array}{ll}
                0 &     x<x_1\\
                \frac{i}{n} & x_i\le x < x_{i+1} \\
                1 & x>x_N \\
                \end{array} \right.
\]

This distribution is compared with the cumulative distribution function $F(x)$
occuring for a Poisson process at the constant rate determined through the 
weighted average. In principle the test is applicable only if the comparison 
function has no parameter determined by the data themselves; that is not our case 
but it is fair to assume that one parameter determined out of a large data 
sample will have little influence on its outcome.\\
The distance between the experimental and theoretical cumulative distributions
is calculated in two different metrics: the first, named after Kolmogorov,
is
\[
D_N = max_{x} |S_N(x)-F(x)|,
\]
the second, named after Smirnov, is
\[
W^2_N = \int_{-\infty}^{+\infty}(S_N(x)-F(x))^2 dF(x). 
\]
The distribution functions of $D_N\sqrt{N}$ and $W^2N$ can be calculated and
are available as analytical formulae, tables or recurrence relations implemented
by library routines.\\
The same approach and the same formulae allows the comparison between two 
set of data to verify if they come from the same original distribution. The 
distance between the cumulative distributions of two samples of size M and N 
is, after Kolmogorov,
\[
D_{NM} = max_{x} |S_N(x)-S_M(x)|
\]
where $D_{NM}\sqrt{MN/(M+N)}$ has the same probability distribution as 
$D_N\sqrt{N}$ and, after Smirnov, 
\[
W^2_{NM} = \int_{-\infty}^{+\infty}(S_N(x)-S_M(x))^2 d\left[\frac{NS_N(X)+
MS_M(X)}{N+M} \right] 
\]
where $W^2_{NM}(MN/(M+N))$ has the same probability distribution as $W_N^2N^2$.
The test is exact because there are no estimated parameters.

\subsection{The full data sample}

The previous hypothesis tests applied to the full data sample under the 
hypothesis of constant flux give the following probabilities: 

\begin{eqnarray*}
P(\chi^2) = 0.33 \% \\
P(D_N) = 3.62 \% \\
P(W^2_N) = 2.79 \% 
\end{eqnarray*}

The smallest one ($\chi^2$) has the limit of relying heavily on a delicate 
procedure of error estimation. The others are more robust but not small enough 
to be conclusive.

\section{Rank measure of association }

Another approach to the data is studying the correlation between the measured
production rate in each run and another run dependent quantity. The candidate 
quantities are the run epoch, the effective run time length and the background
rate. Alternatively any quantity defined versus the time epoch can be used. 
In \cite{bapr}, \cite{krauss}, \cite{bieber} several sun's activity dependent
quantities are used, e.g. the mean sunspot number.\\
An estimation of correlation can be done using the standard correlation 
coefficient as in \cite{krauss} according to \cite{press} or using rank-order 
statistic as in \cite{bapr}. The latter is more robust because it makes no
assumption on the underlying distributions nor on the data errors. As suggested
in \cite{bapr} and described in \cite{press} the rank statistical tests of 
Spearman rank-order correlation coefficient and Kendall's $\tau$ are applied.\\
The principle is that if two quantities labelled by run index are rank ordered, 
the more uniform is the resulting scatter plot the less the two quantities are 
correlated.\\ 
In Fig.\ref{countbacka}, the background rate is plotted versus the $^{37}Ar$ 
production rate, as well as the corresponding rank ordered quantities. The rank 
ordered plot shows a denser diagonal band suggesting a significant 
anticorrelation.
 
\subsection{The full data sample}

The Spearman and Kendall tests are applied to the full data sample measuring the
correlation between the $^{37}Ar$ production rate and the run epoch ($t_r$), the 
effective run time length ($t_{eff}$) and the background rate (back). The result
is expressed as the probability to obtain the observed correlation parameter 
from two uncorrelated distributions: 

{
\begin{tabular}{l l l l}
                         & $t_{eff}$ & $t_{r}$ & back \\
{ Spearman $r_s$} & 4.95\% & 2.09\% & $9.5\, 10^{-3}$\% \\
{ Spearman D}  & 5.15\% & 2.04\% & $14.3\, 10^{-3}$\% \\
{ Kendall}     & 5.76\% & 1.88\% & $14.9\, 10^{-3}$\% \\

\end{tabular}
}

\begin{figure*}
\begin{center}
\mbox{\begin{tabular}[t]{ll}
\subfigure[]{\includegraphics[width=9cm]{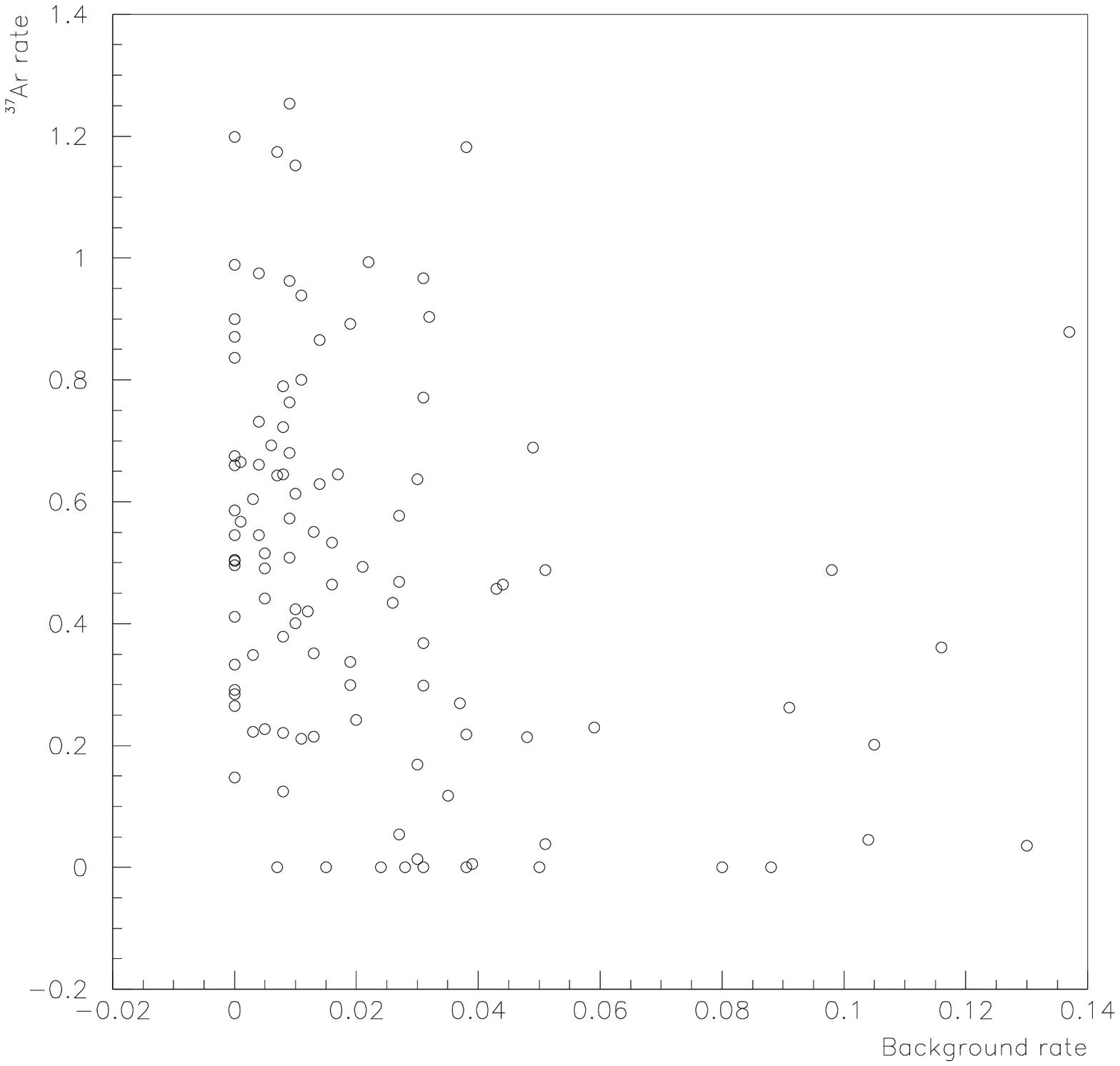}
\label{countbacka}
} &
\subfigure[]{\includegraphics[width=9cm]{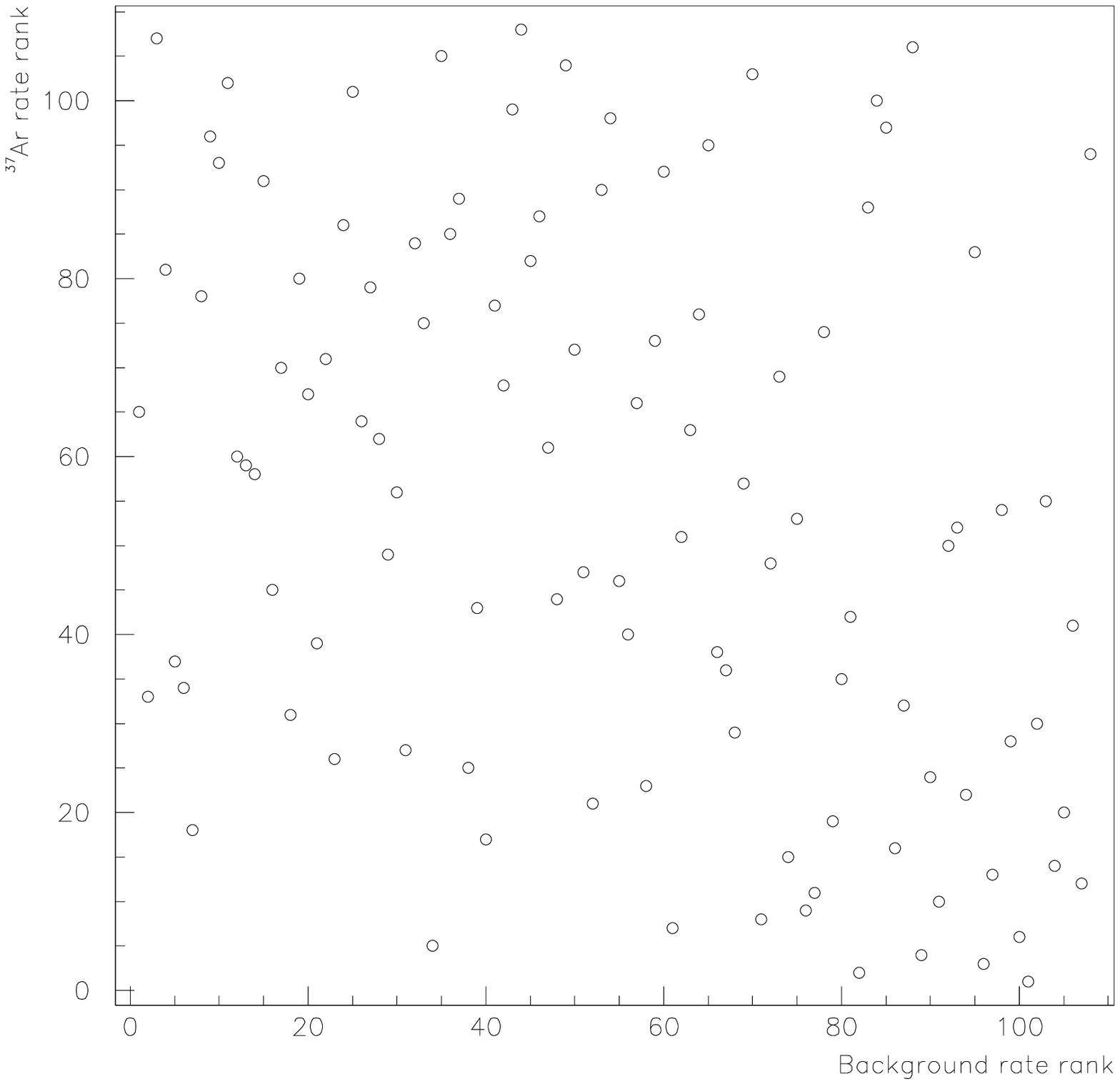}
\label{countbackb}
} \\
\end{tabular}}
\caption{a) Background rate versus $^{37}Ar$ production rate b) Background rate 
rank versus $^{37}Ar$ production rate rank. In the null hypothesis of no 
correlation, the points should be uniformly distributed.}
\end{center}
\end{figure*}

From this table it is apparent that there is little (anti)-correlation between the
Argon production rate and $t_{eff}$, a little more with the run epoch and a very
strong one with the background level, that is visible in Fig.\ref{countbackb}.
This correlation has no physical justification and has never been noticed 
before. Its strenght is comparable to the strongest correlation identified in 
previous papers \cite{krauss}, \cite{bapr}, \cite{bieber}.\\
It suggests that the analysis algorithm decreases the signal in presence of 
background and therefore that the average production rate is underestimated.

\section{Partitioning the data set }

\begin{figure*}
\begin{center}
\mbox{\begin{tabular}[t]{ll}
\subfigure[]{\includegraphics[width=9cm]{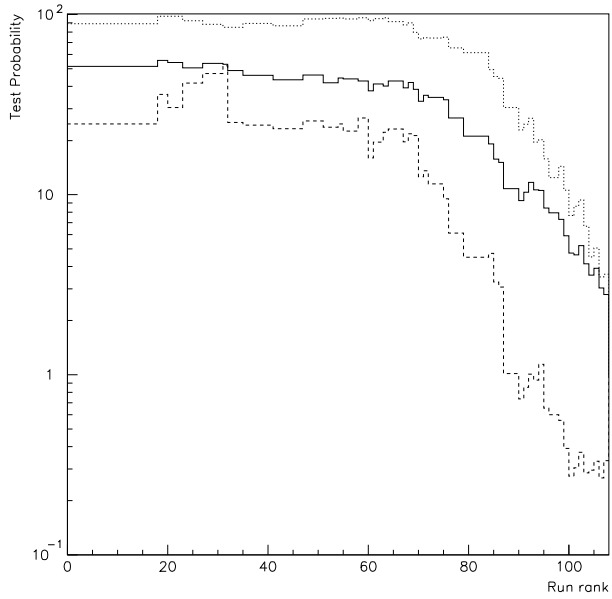}
\label{diffbacka}
} &
\subfigure[]{\includegraphics[width=9cm]{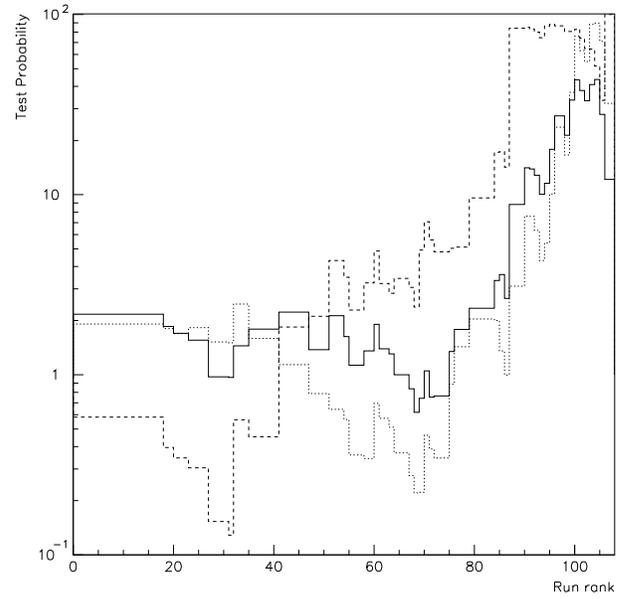}
\label{diffbackb}
} \\
\subfigure[]{\includegraphics[width=9cm]{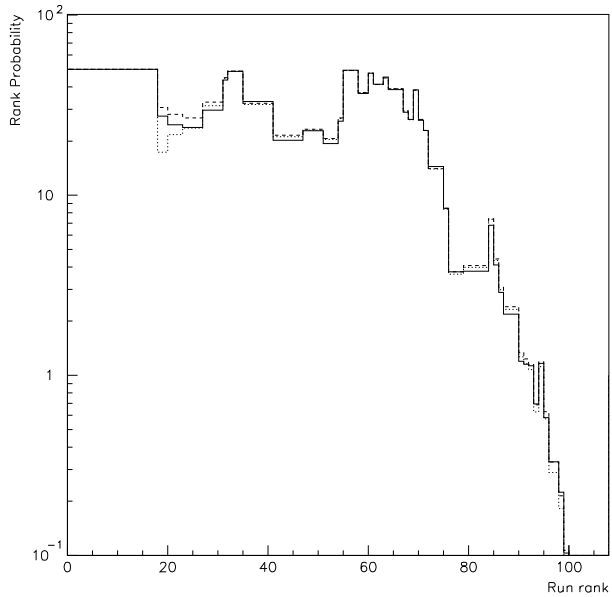}
\label{diffbackc}
} &
\subfigure[]{\includegraphics[width=9cm]{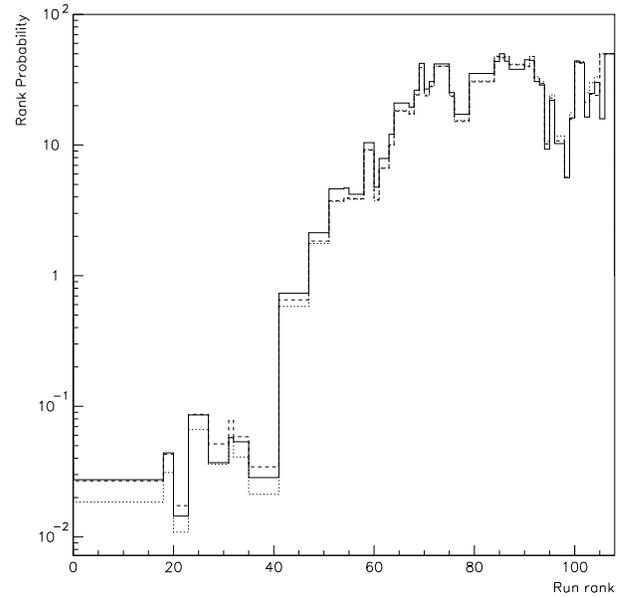}
\label{diffbackd}
} \\
\end{tabular}}
\caption{
a) Probability of constant $^{37}Ar$ rate of the lowest N background runs
for Smirnov (solid), Pearson (dashed) and Kolmogorov (dotted)
b) Probability of constant $^{37}Ar$ rate of the highest $N_T-N$ background runs
for Smirnov (solid), Pearson (dashed) and Kolmogorov (dotted)
c) Probability of no correlation between $^{37}Ar$ rate and background of the 
lowest N background runs for Kendall (solid), Spearman D (dashed) and Spearman 
$r_s$ (dotted)
d) Probability of no correlation between $^{37}Ar$ rate and background of the 
highest $N_T-N$ background runs for Kendall (solid), Spearman D (dashed) and 
Spearman $r_s$ (dotted)
 .}
\end{center}
\end{figure*}

\begin{figure*}
\centering
\includegraphics[width=\textwidth,bb=0 280 567 567]{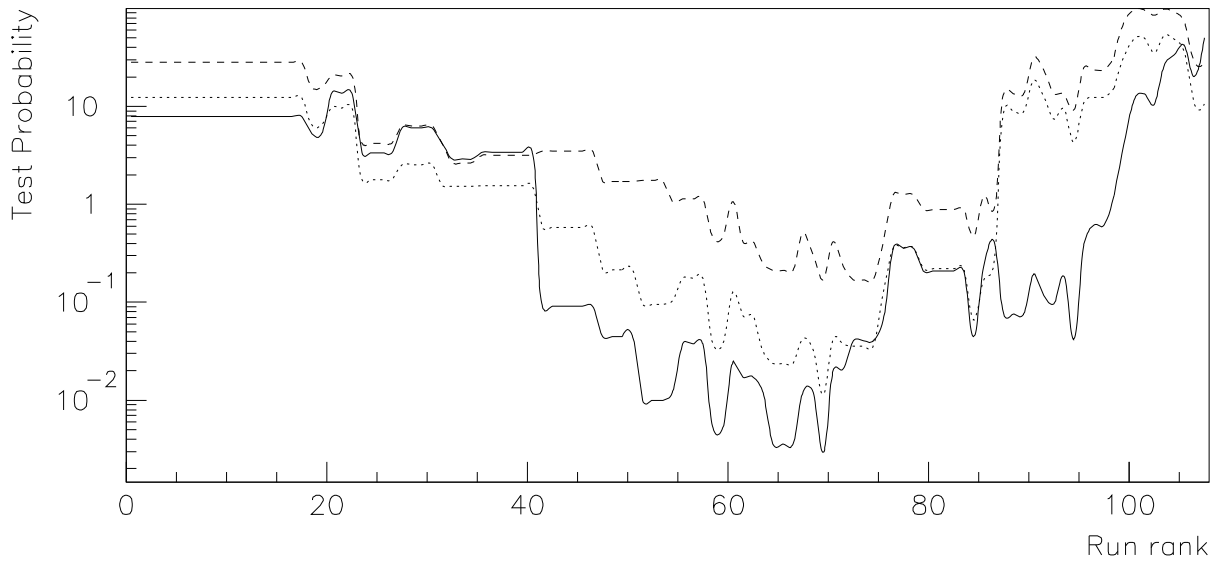}
\caption{Probability of constant and equal $^{37}Ar$ rate between the lowest 
N and highest $N_T-N$ background runs
for Smirnov (solid), Pearson (dashed) and Kolmogorov (dotted) }
\label{bpdiff}
\end{figure*}

\begin{figure*}
\centering
\includegraphics[width=\textwidth,bb=0 280 567 567]{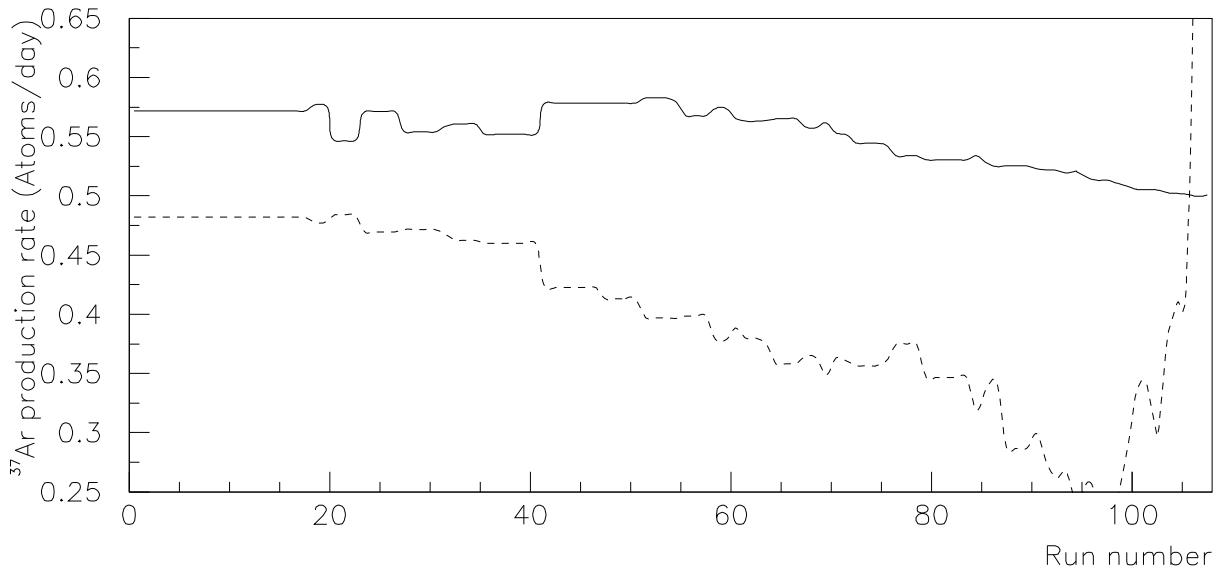}
\caption{Weighted average of the $^{37}Ar$ rate of the lowest 
N (solid) and highest $N_T-N$ (dashed) background runs }
\label{bave}
\end{figure*}

The previous results provide strong suggestions that the $^{37}Ar$ production 
rate is anticorrelated with the background and that the overall consistency of
the data under the hypothesis of constant flux is poor. Partitioning the data 
set in two subsets might help to gain more understanding on this disagreement.\\
To limit the arbitrariness of the partition, we consider the run set ranked 
according to the background rate and, for each number $1\le N<N_T$, the 
partition in the two sets including the runs with the $N$ lower background rate 
and the runs with the $N_T-N$ higher background rate. That gives $N_T-1$ 
partitions in two sets.\\
For each set in each partition the average production rate is recalculated and 
the previous hypothesis tests are applied as well as the rank order tests. The 
hypothesis tests are applied also on the pair of sets of each partition
to estimate the probablity that they come from the same constant distribution.\\
In Fig.\ref{diffbacka}-\ref{diffbackb}-\ref{diffbackc}-\ref{diffbackd} 
the test and rank probabilities are plotted versus the
background rate ranked run number. In Fig.\ref{bpdiff} the probabilities of
low and high background ranked runs coming from the same constant distribution
are plotted and in Fig.\ref{bave} the average $^{37}Ar$ production rate 
for the two sets is shown.\\
What is apparent is that when the hypothesis and the rank tests are restricted 
to the lower two third of the runs (about 70), the experimental data are 
coherent with the hypothesis of constant $^{37}Ar$ production rate and the
average value is constant with the run rank cut. Also the upper third of the 
runs, albeit less clearly, is a data set coherent with the hypothesis of 
constant $^{37}Ar$ production rate even if the average $^{37}Ar$ production 
rate depends on the run rank cut.\\
The plot in Fig.\ref{bpdiff} demonstrate that the probability that the two
complementary sets belong to the same distributions reaches a minimum around a
value of 70.\\
The most natural interpretation is that the low and high background sets come
from two distinct populations with different averages. The low background
population is highly coherent and unbiased and therefore gives the most 
reliable estimate of the $^{37}Ar$ production rate. The high backgorund 
population is less coherent, as it is to be expected if the measurement of
the production rate is biased by the background, and does not provide a clear
measurement of the average.

\section{Conclusion}

The conclusion from the previous results is that the analysis on the data from
the Homestake experiment should be restricted to the subset of about two third
of the runs with low background. The runs with large background should be
discarded.\\
The somehow arbitrary choice of the cut in the backgorund rank adds a small
uncertainties to the estimation of the $^{37}Ar$ production rate. Choosing 
$N=70$, using the weighted average as estimator and retaining the same 
systematic error of the original paper, the result is
\[
0.566 \pm 0.030 (statistical) \pm 0.030 (systematic) day^{-1}
\]
or
\[
3.03 \pm 0.16 (statistical) \pm 0.16 (systematic) SNU
\]
That is larger of almost three statistical standard deviation than the original
result in \cite{homestake}

\bibliographystyle{apj}
\bibliography{homestake}

\end{document}